\documentclass[%
 reprint,
 amsmath,amssymb,
 aps,
pra,
]{revtex4-2}

\usepackage{graphicx}
\usepackage{dcolumn}
\usepackage{bm}
\usepackage{lipsum}
\usepackage{xcolor}
\usepackage{amsmath}
\begin{document}
\title{Born-Padé approach to electromagnetic scattering in complex one-dimensional inhomogeneous slabs}

\author{J. A. Rebouças}
    \email{jalvesreboucas@gmail.com}
    \affiliation{Instituto de Educação Ciência e Tecnologia do Ceará, Iguatu, CE, Brazil} 
    \affiliation{Universidade Federal de Alagoas, Maceió, AL, Brazil}
    
\author{P. A. Brandão}
    \email{paulo.brandao@fis.ufal.br}
    \affiliation{Universidade Federal de Alagoas, Maceió, AL, Brazil}

\date{\today} 

\begin{abstract}
Perturbation theory is applied to one-dimensional scattering systems consisting of a general class of inhomogeneous and isotropic slabs having size $L$ described by the relative permittivity $\varepsilon(z) = 1 + \alpha \chi(z)$, where $\chi(z)$ is the electric susceptibility and $\alpha$ the perturbation parameter. The transmitted and reflected amplitudes are shown to be written as Born series in powers of $\alpha$ and Padé approximants are used to obtain analytical results with a high and arbitrary degree of accuracy. The approach is very general and can handle oblique incidence. Examples are given for the transmission and reflection amplitudes of plane waves interacting with Hermitian and non-Hermitian structures with known and unknown exact solutions.
\end{abstract}

\maketitle

\section{Introduction}
A very important theoretical problem in optics is the exact and analytical determination of the scattering amplitudes, for the transmitted and reflected plane waves, after interacting with an inhomogeneous material described by the relative dielectric permittivity $\varepsilon(z)$ and occupying a finite extension of size $L$. If the slab is composed of several dielectric layers, with constant but distinct values of $\varepsilon$ in each layer, there are well-known procedures, such as the transfer and scattering matrix formalism, that can be used to obtain exact results, usually by performing numerical operations involving matrix multiplication \cite{born2013principles,saleh2019fundamentals}. By using matrices to obtain the scattered amplitudes, one has the obvious advantage in that it can be easily implemented in computers and give fast results. The disadvantage is that one loses track of what is going on during this process, especially if there are a large number of layers, and the procedure will ultimately depend on the numerical algorithm which behaves like a black box without a more intuitive feeling about the process. Moreover, if the material is intrinsically inhomogeneous, this method can still be used but at the expense of introducing numerical errors arising from approximating the continuous material by piecewise constant parts.

Another method that is generally valid in scattering systems is perturbation theory \cite{bender1999advanced}. Perturbation methods substitute a very hard problem into a sequence of (presumably) much easier ones. The approach taken to obtain the scattered amplitudes in the majority of cases, especially in two or three dimensions, is almost always dependent on the Born approximation, which is used under the assumption of a weak scatterer. The Born approximation is obtained from perturbation theory by rejecting all terms in the Taylor expansion of the scattered field except the first non-trivial one. Unfortunately, if the objective is to go beyond the Born approximation, in most cases the perturbation series diverges in the formal sense. Since it is highly desirable to derive results under regimes where the scatterer strongly interacts with the incident field, there are two main routes to take from here: (1) To choose another formulation, relying on heavily numerical computations or (2) to sum the divergent series. This paper is based on the second route applied to one-dimensional scattering involving inhomogeneous isotropic materials.

Analytic or semi-analytic scattering theories for plane waves interacting with inhomogeneous slabs were considered before \cite{bremmer1951wkb,atkinson1960wave,berk1967convergence,hassab1972perturbational,chen1978integral,chen1979finite,chen1980variational,su1982fast}. Particular importance was given to the work of Bremmer \cite{bremmer1951wkb} who studied this problem from a very physical point of view by discretizing the slab, calculating the refractions through the material, and then performing the continuous limit to obtain an analytical solution to the problem. The solution obtained by Bremmer has the form of a series whose first term is the WKB approximation \cite{bremmer1951wkb,atkinson1960wave}. The convergence of Bremmer series was also addressed \cite{berk1967convergence}.

One-dimensional inhomogeneous slabs have found numerous recent applications in optics due especially to the emergence of non-Hermitian photonics \cite{longhi2018parity,el2019dawn,feng2017non}. The fact that a quantum system described by a non-Hermitian Hamiltonian having Parity-Time (PT) symmetry can give real eigenvalues \cite{bender1998real} has a direct consequence in optical-analog systems. Effects such as lasing modes in active cavities \cite{ge2011unconventional}, unidirectional invisibility \cite{lin2011unidirectional,jones2012analytic}, slabs with random properties \cite{kalish2012light}, anisotropic reflection in Bragg systems \cite{zhu2014anisotropic} were considered in this context. Thus, the development of analytical methods are useful in that one can explore a more general class of inhomogeneous materials. 

This paper is organized as follows: In Section II(A) we develop the Born series for the scattering problem by using perturbation theory. Section II(B) introduces the Pad\'e approximants for the reflected and transmitted amplitudes. Section III is devoted to applications. In part (A) we consider a homogenous slab, part (B) a linear slab, part (C) a periodic slab with PT symmetry and in part (D) a slab without known closed-form analytic solution consisting of a combination of the linear and periodic cases.

\section{Theory}

This section is composed of two subsections. The first one deals with the development of the Born series through the use of perturbation theory. The second subsection introduces the idea of a Pad\'e approximant to represent the transmitted and reflected amplitudes of the scattered waves.

\subsection{Born series for the Helmholtz equation}

A monochromatic component $E(x,z;\omega)$ of the electric field polarized in the $y$ direction satisfies the Helmholtz equation
\begin{equation}
     \frac{\partial^2 E}{\partial z^2} + \frac{\partial^2 E}{\partial x^2} + k^2\varepsilon(z)E = 0,
\end{equation}
where $k = \omega/c$, with $c$ being the speed of light in vacuum, $\omega$ the angular frequency, $\varepsilon(z) = 1 + \alpha\chi(z)$ is the inhomogenenous dielectric constant, $\chi(z)$ is the dielectric susceptibility and $\alpha$ the expansion parameter that we put equal one at the end of the calculations. We assume that the material is confined in the region between $0$ and $L>0$ and write $\chi(z) = \Theta_L(z)\xi(z)$, where $\Theta_L(z) = 1$ if $z \in [0, L]$ and $\Theta_L(z) = 0$ if $z \notin [0,L]$. We leave $\xi(z)$ unspecified for the moment but it can be a real or complex function of $z$. The real electric field is obtained from $\mathbf{E}(x,z,t) = \text{Re}[E(x,y;\omega)e^{-i\omega t}]\hat{y}$.

\begin{figure}[!htp]
    \centering
    \includegraphics[scale=0.2]{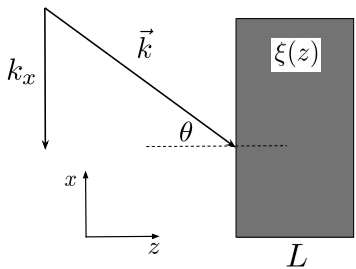}
    \caption{Scattering geometry. The incident plane wave is characterized by the wavevector $\mathbf{k}$ which makes an angle $\theta$ with the normal to the interface located at $ z = 0$. The parallel component $k_x$ to the dielectric interface is given by $k_x = -k\sin\theta$. }
    \label{fig1}
\end{figure}

To take into account effects arising from the incident angle of the plane wave, the field is written as (see Figure \ref{fig1})
\begin{equation}
    E(x,z) = \psi(z)e^{ik_x x},
\end{equation}
where $\psi(z)$ is a $z$-dependent field amplitude and $k_x$ is the wavevector component perpendicular to the propagation direction. If $\theta \in (-\frac{\pi}{2},\frac{\pi}{2})$ is the angle formed between the incident wavevector $\mathbf{k}$ and the $z$ axis, then $k_x = -k\sin\theta$ and the differential equation satisfied by $\psi(z)$ can be written as
\begin{equation}\label{eq3}
    \frac{d^2\psi(z)}{dz^2} + k^2\psi(z)\cos^2\theta = -k^2\alpha\chi(z)\psi(z).
\end{equation}
The Green's function method can be used to cast Eq. \eqref{eq3} into an integral relation. The Green's function $G(z,z')$ for the slab-free Helmholtz equation is obtained in the usual way. The result is given by
 \begin{equation}
     G(z,z') = \frac{e^{ik\cos\theta|z-z'|}}{2ik\cos\theta},
 \end{equation}
and \eqref{eq3} can be rewritten as
\begin{equation}
\begin{split}
    \psi(z) &= e^{ikz\cos\theta} - \alpha k^2\int_{-\infty}^{+\infty} dz' G(z,z')\chi(z')\psi(z') \\
    &= e^{ikz\cos\theta} - \alpha k^2\int_{0}^{L} dz' G(z,z')\xi(z')\psi(z'),
\end{split}
\end{equation}
where the first term on the right-hand side is a unit amplitude plane wave which is the solution to the homogeneous problem with $\alpha = 0$ (no scatterer).

To proceed, the field amplitude $\psi(z)$ is assumed to be written as a series expansion in powers of $\alpha$:
\begin{equation}
    \psi(z) = \sum_{n = 0}^{\infty}\psi_n(z)\alpha^n,
\end{equation}
with $\psi_0(z) = e^{ikz\cos\theta}$ being the solution to the unperturbed problem with $\alpha = 0$. The remaining coefficients $\psi_n$ $(n\geq 1)$ are related recursively by
\begin{equation}
    \psi_n(z) = -k^2\int_0^L dz'G(z,z')\xi(z')\psi_{n-1}(z').
\end{equation}
Thus, to obtain the reflected and transmitted fields it is necessary to solve the integral inside the scatterer. Direct use of the Green's function and the incident field $u_0$ allows us to write
\begin{equation}
    \begin{split}
        \psi_n(z) &= P_n(z)e^{ikz\cos\theta} + N_n(z)e^{-ikz\cos\theta},
    \end{split}
\end{equation}
valid in the region $0 \leq z \leq L$, where $P_n$ and $N_n$ are determined recursively through the system of equations
\begin{equation}\label{systemR}
    \begin{split}
        P_n(z) &= \frac{ik}{2\cos\theta}\int_0^z ds \xi(s)P_{n-1}(s)\\
        &+ \frac{ik}{2\cos\theta} \int_0^z ds \xi(s)N_{n-1}(s)e^{-2iks\cos\theta},
    \end{split}
\end{equation}
\begin{equation}
    \begin{split}\label{systemL}
        N_n(z) &= \frac{ik}{2\cos\theta}\int_z^L ds \xi(s)P_{n-1}(s)e^{2iks\cos\theta}\\
        &+ \frac{ik}{2\cos\theta} \int_z^L ds \xi(s)N_{n-1}(s),
    \end{split}
\end{equation}
with the initial conditions $P_0 = 1$ and $N_0 = 0$. 

Finally, the transmitted field in the region $z > L$ can be calculated by using
\begin{equation}
    \psi_n(z) = t_ne^{ikz\cos\theta} \quad (z > L), 
\end{equation}
where $t_n$ is given by
\begin{equation}\label{tn}
\begin{split}
    t_n = \frac{ik}{2\cos\theta}&\Bigg[\int_0^L ds\xi(s)P_{n-1}(s) \\
    &+  \int_0^L ds\xi(s)N_{n-1}(s)e^{-2iks\cos\theta} \Bigg]
\end{split}
\end{equation}
and the reflected field in the region $z < 0$ can be calculated from
\begin{equation}
    \psi_n(z) = r_ne^{-ikz\cos\theta} \quad (z < 0),
\end{equation}
with $r_n$ given by
\begin{equation}\label{rn}
    \begin{split}
        r_n = \frac{ik}{2\cos\theta}&\Bigg[\int_0^L ds\xi(s)P_{n-1}(s)e^{2iks\cos\theta} \\
    &+  \int_0^L ds\xi(s)N_{n-1}(s) \Bigg].
    \end{split}
\end{equation}

For a given material distribution $\xi(z)$, Eqs. \eqref{systemR} and \eqref{systemL} can be solved by iteration and the coefficients $t_n$ and $r_n$ are calculated by using Eqs. \eqref{tn} and \eqref{rn}. The total transmitted $t$ and reflected $r$ amplitudes are given by
\begin{equation}\label{trTaylor}
    \begin{split}
        t &= \sum_{n= 0}^{\infty}t_n\alpha^n = 1 + t_1\alpha + t_2\alpha^2 + ...,\\
        r &= \sum_{n= 0}^{\infty}r_n\alpha^n = 0 + r_1\alpha + r_2\alpha^2 + ...,
    \end{split}
\end{equation}
where $r_0 = 0$ and $t_0 = 1$ are the solutions to the unperturbed problem (wave is fully transmitted) and $(t_1,r_1)$ are the first-order Born approximations for the transmitted and reflected amplitudes. By the end of the analysis we substitute $\alpha = 1$ to obtain the scattering amplitudes for the original slab with $\varepsilon(z) = 1 +\chi(z)$.

\subsection{Padé approximants}

By following the recipe presented in the previous section, one obtains the transmitted and reflected amplitudes represented by a Taylor series expansion in powers of $\alpha$. In most cases of interest, however, the series will be of a divergent type when $\alpha = 1$ is substituted in the expressions, invalidating the equality between $t$($r$) and $\sum_{n}t_n\alpha^n$ ($\sum_n r_n\alpha^n$). Therefore, it is important to ask if there is another representation available that yields accurate and convergent results in a larger region in the complex $\alpha$-plane.

One such representation seems to handle these issues very well. The Padé approximants $P_N^M$ are a family of rational functions whose denominator and numerator are polynomials of degree $N$ and $M$, respectively:
\begin{equation}
    P_N^M = \frac{\sum_{n = 0}^M A_n\alpha^n}{\sum_{n = 0}^N B_n \alpha^n} = \frac{A_0 + A_1\alpha+A_2\alpha^2 + ... + A_M\alpha^M}{1 + B_1\alpha + ... + B_N\alpha^N},
\end{equation}
where $A_n$ and $B_n$ are coefficients and $B_0 = 1$ is assumed without loss of generality \cite{bender1999advanced,baker1996pade,brezinski2012history,george1975essentials}. Given a Taylor series $\sum_{n=0}^{\infty}a_n\alpha^n$, the idea is to match the first $N + M + 1$ coefficients in the Taylor expansion of the Pad\'e representations with that of its corresponding Taylor series. For example, the first diagonal $(N=M=1)$ Padé approximant $P_1^1$ is given explicitly by
\begin{equation}\label{pade11}
        P_1^1 = \frac{A_0 + A_1\alpha}{1 + B_1\alpha} = a_0 + \frac{a_1^2\alpha}{a_1 - a_2\alpha}.
\end{equation}
In the particular case of diagonal Pad\'e approximants ($N = M$), the representations can be obtained directly by performing the division between the determinant of two $(N+1)\times(N+1)$ matrices $\mathbf{Q}$ and $\mathbf{L}$, given by
\begin{equation}
\begin{split}
\mathbf{Q} &= 
    \begin{pmatrix}
    a_1 & a_2 & \cdots & a_{N+1} \\ a_2 & a_3 & \cdots & a_{N+2} \\ \vdots & \vdots & \ddots & \vdots \\ a_N & a_{N+1} & \cdots & a_{2N} \\ a_0\alpha^N & a_0\alpha^{N-1}+a_1\alpha^N & \cdots & \sum_{j=0}^N a_j\alpha^j
    \end{pmatrix},\\
\mathbf{L} &= 
    \begin{pmatrix}
    a_1 & a_2 & \cdots & a_{N+1} \\ a_2 & a_3 & \cdots & a_{N+2} \\ \vdots & \vdots & \ddots & \vdots \\ a_N & a_{N+1} & \cdots & a_{2N} \\ \alpha^N & \alpha^{N-1} & \cdots & 1
    \end{pmatrix}.
    \end{split}
\end{equation}
Thus, we can identify $P_N^N = \text{det}\mathbf{Q}/\text{det}\mathbf{L}$. More general matrices can be viewed in Ref. \cite{baker1996pade}. We hope that the Padé approximants become closer to the exact answer as $N,M\rightarrow\infty$. In what follows, we consider only the diagonal approximants with $N = M$. The transmitted $t$ and reflected $r$ amplitudes have Pad\'e representations given by
\begin{equation}
    t_M^M = \frac{\sum_{n = 0}^{M}\mathcal{A}_n\alpha^n}{1 + \sum_{n = 1}^{M}\mathcal{B}_n\alpha^n},
\end{equation}
\begin{equation}
    r_M^M = \frac{\sum_{n = 0}^{M}\mathcal{C}_n\alpha^n}{1 + \sum_{n = 1}^{M}\mathcal{D}_n\alpha^n}.
\end{equation}
There are well-known algorithms which relates $\mathcal{A}_n$ and $\mathcal{B}_n$ $(\mathcal{C}_n$ and $\mathcal{D}_n)$ with $t_n$ $(r_n)$ \cite{bender1999advanced,baker1996pade,brezinski2012history,george1975essentials}.

The most remarkable property of Padé approximants is that they are able to represent a given function in a larger region in the complex $\alpha$-plane, when compared to the convergence region in the Taylor representation of the same function. In other words, the approximants approach a limit even if the Taylor representation diverges. Moreover, they converge faster than Taylor approximants (partial sums) in most cases. It is also verified that only very few initial terms in the Taylor expansion must be known to form the first approximants, and this represents a major advantage since one is usually not in possession of all the terms in a general perturbation problem. More explicitly, the first $2N+1$ coefficients in the Taylor series are necessary to form the approximant $P_N^N$. 

Unfortunately, there are no general theorems on the questions of existence and uniqueness of the approximants to arbitrary functions with the exception of Stieltjes functions. Nevertheless, there is a large amount of evidence that the approximants can handle a variety of functions other than Stieltjes. Research in this direction is still ongoing.

\section{Applications}

To validate the above formalism and the Pad\'e representations for the transmitted and reflected amplitudes, we consider a few applications involving dielectric systems with known and unknown exact solutions. Recent works on the use of Pad\'e approximants in classical optics can be found in Refs. \cite{rebouccas2021scattering, van2020electromagnetic}.

\subsection{Hermitian and non-Hermitian homogeneous dielectric slab}

The first example we consider is the most simple one where $\xi(z) = b$ with $b$ a real or complex number, independent of $z$. The scatterer is a homogeneous dielectric slab of size $L$. In this case, the first two coefficients $P_1$ and $N_1$, obtained directly from Eqs. \eqref{systemR} and \eqref{systemL}, are given by 
\begin{equation}
\begin{split}
    P_1(z) &= \frac{ibkz}{2\cos\theta}, \\ N_1(z) &= b\frac{e^{2ikL\cos\theta} - e^{2ikz\cos\theta}}{4\cos^2\theta}.
    \end{split}
\end{equation}
From Eqs. \eqref{tn} and \eqref{rn} we obtain the first two corrections to the transmitted and reflected amplitudes,
\begin{equation}\label{firstts}
\begin{split}
    t_1 &= \frac{ikbL}{2\cos\theta}, \\
    t_2 &= \frac{ikb^2}{8\cos^2\theta}\Bigg[ ikL^2 - \frac{L}{\cos\theta} + \frac{(e^{2ikL\cos\theta} - 1)}{2ik\cos^2\theta} \Bigg],
\end{split}
\end{equation}
and
\begin{equation}\label{firstrs}
    \begin{split}
        r_1 &= \frac{b}{4\cos^2\theta}\Big( e^{2ikL\cos\theta} - 1 \Big),\\
        r_2 &= \frac{b^2}{8\cos^4\theta}\Big[ e^{2ikL\cos\theta}(2ikL\cos\theta-1)+1\Big],
    \end{split}
\end{equation}
which represent the first and second Born approximations, respectively. This iterative procedure can be carried out until one reaches a desired approximation order. Explicitly, with $(t_0,t_1,t_2)$ and $(r_0,r_1,r_2)$ given above, the first-order Pad\'e approximant can be readily obtained by using Eq. \eqref{pade11}. More coefficients are necessary to construct high-order approximants.

\begin{figure}[!htp]
    \centering
    \includegraphics[scale=0.67]{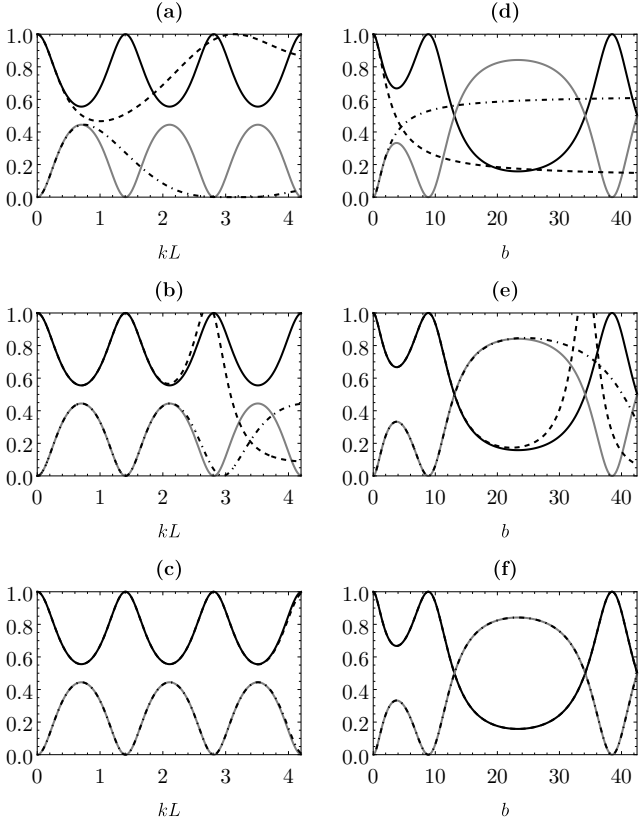}
   \caption{\label{fig2} Transmission and reflection amplitudes for a homogeneous slab with $\xi (z) = b$, independent of $z$. The continuous black line is the exact (absolute squared) solution for the transmitted amplitude and the continuous gray line is the exact solution for the (absolute squared) reflected amplitude. The dashed (dashed-dotted) lines are the Pad\'e approximants for the transmitted (reflected) amplitudes (a,d) $|t_1^1|^2$ and $|r_1^1|^2$, (b,e) $|t_3^3|^2$ and $|r_3^3|^2$, (c,f) $|t_5^5|^2$ and $|r_5^5|^2$. Parameters used: $\alpha = 1$, $\theta = 0$, (a-c)$b = 4$ and (d-e)$kL = 1$.}
    \label{fig1p2}
\end{figure}

We present in Figure \ref{fig1p2} several plots of the approximants $|t_N^N|^2$ and $|r_N^N|^2$ for $N = 1$, 3 and 5 along with the exact solution (see Appendix A). First of all, since the exact solution for the scattered amplitudes is not a quotient of polynomials in $\alpha$, the Pad\'e approximants can never return the exact answer in this case. Nevertheless, it is remarkable how they are able to adapt to the exact solution as $N$ increases. In parts (a-c) of Figure \ref{fig2} we display the scattering amplitudes as a function of $kL$. The approximation becomes better as $N$ increases, as expected, and for $N = 5$ we already obtain a very accurate result for the range of parameters used. Notice that to calculate $t_5^5$ and $r_5^5$ only 11 terms in the Taylor expansion are used. Parts (d-f) display the behavior of the scattering amplitudes as a function of $b$. This case is more remarkable because $t_n \sim b^n$ and $r_n \sim b^n$ (which is expected since for a homogeneous slab we can take $b$ as the perturbation parameter) so the Taylor representation is certainly to be of a divergent type, especially for values of $b$ as large as 40. Indeed, a plot of the partial sums of the power series in $\alpha$ having the same number of terms as in the corresponding Pad\'e representations diverges \textit{wildly} (not shown). Nevertheless, the Pad\'e approximants are able to recover the true behavior. We remark again that there are no numerical errors associated with these results in the sense that they represent analytic approximations.

To see how the Pad\'e's handle gain and loss, Figure \ref{fig3} shows the plot of $|t_N^N|^2$ and $|r_N^N|^2$ as a function of $kL$ for homogeneous slabs with loss and and gain, i.e, $\xi(z) = 4 \pm i$. Clearly, the approximants have no trouble in dealing with a lossy/active layer.

\begin{figure}[!htp]
    \centering
    \includegraphics[scale=0.67]{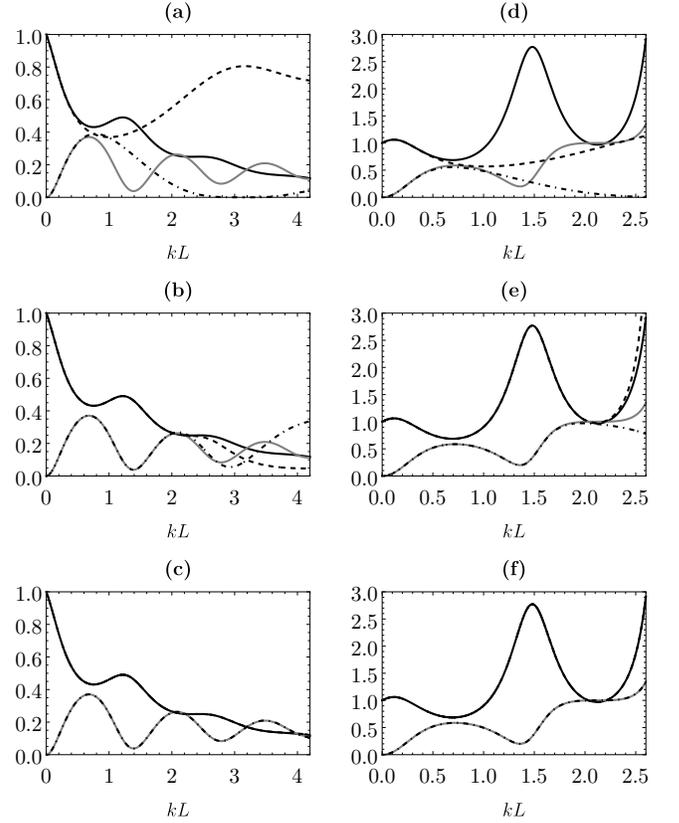}
   \caption{\label{fig3} Transmission and reflection amplitudes for a homogeneous slab with (a-c) $\xi (z) = 4+i$ and (d-f) $\xi(z) = 4 - i$ as a function of $kL$. The continuous black line is the exact (absolute squared) solution for the transmitted amplitude and the continuous gray line is the exact solution for the (absolute squared) reflected amplitude. The dashed (dashed-dotted) lines are the Pad\'e approximants for the transmitted (reflected) amplitudes. (a,d) $|t_1^1|^2$ and $|r_1^1|^2$, (b,c) $|t_3^3|^2$ and $|r_3^3|^2$ and (c,f) $|t_5^5|^2$ and $|r_5^5|^2$. Parameters used: $\theta = 0$ and $\alpha = 1$}
    \label{fig3}
\end{figure}

\subsection{Linear slab}
Another interesting class of dielectrics are the linear materials described by the profile $\xi(z) = 1 + gz$, with $g$ being a constant parameter. The Helmholtz equation for this case can be recast into Airy's equation of the form $d^2\psi_2(u)/du^2 = u\psi_2(u)$ and we provide in Appendix B the exact solution. By following the same recipe as in the previous section, one obtains the approximants to a desired order of accuracy.

\begin{figure}[!htp]
    \centering
    \includegraphics[scale=0.67]{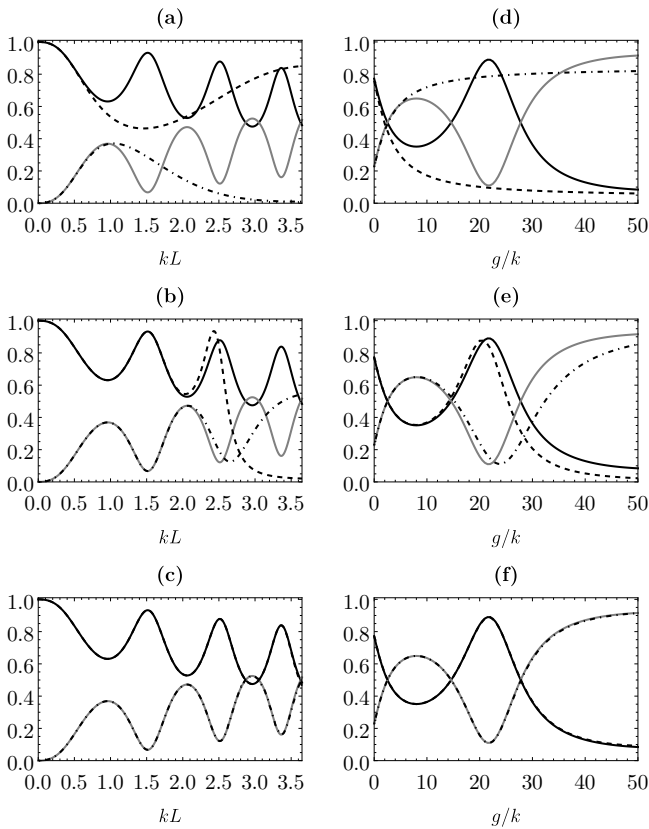}
   \caption{ Transmission and reflection amplitudes for a linear slab with $\xi (z) = 1+ gz$ as a function of (a-c) $kL$ and of (d-f) $g/k$. The continuous black line is the exact (absolute squared) solution for the transmitted amplitude and the continuous gray line is the exact solution for the (absolute squared) reflected amplitude. The dashed (dashed-dotted) lines are the Pad\'e approximants for the transmitted (reflected) amplitudes. (a,d) $|t_1^1|^2$ and $|r_1^1|^2$, (b,f) $|t_3^3|^2$ and $|r_3^3|^2$, (c) $|t_6^6|^2$ and $|r_6^6|^2$ and (e) $|t_2^2|^2$ and $|r_2^2|^2$. Parameters used: $\alpha = 1$, (a-c) $\theta = 0$, $g = 4$, (d-f) $kL = 1$ and $\theta = \frac{\pi}{4}.$}
    \label{fig4}
\end{figure}

The comparison between the exact scattering amplitudes and their Pad\'e representations is shown in parts (a-c) of Figure \ref{fig4} as $kL$ varies. Again, the Pad\'e's are able to approximate the exact answer with an excellent precision, even for an oblique incident plane wave with $\theta = \frac{\pi}{4}$. In parts (d-f) we plot the transmission and reflection as function of $g$ for $g$-values as large as 50.

\subsection{Periodic slab having $PT$ symmetry at the symmetry-breaking point}

\begin{figure}[!htp]
    \centering
    \includegraphics[scale=0.67]{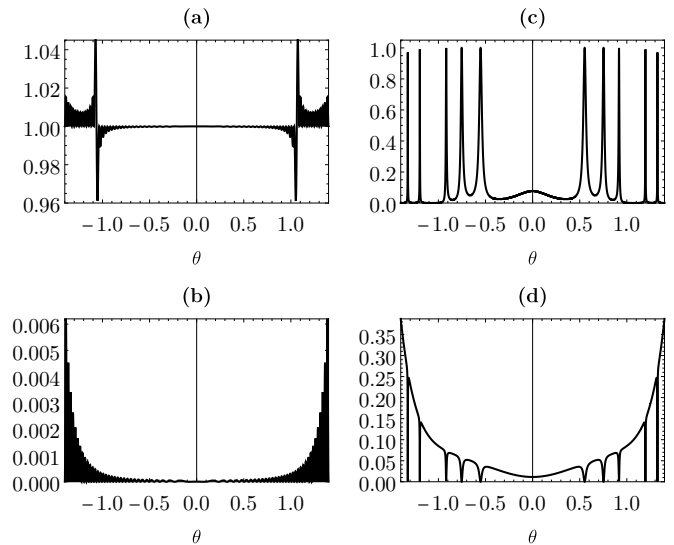}
   \caption{Transmitted and reflected amplitudes for a periodic slab with $\xi (z) = b e^{idz}$ as a function of $\theta$. (a) $|t_2^2|^2$, (b) $|r_2^2|^2$, (c) $|t_4^4|^2$ and (d) $|r_4^4|^2$. Parameters used: $\alpha = 1$, $d = \frac{2\pi}{0.42}$, $k = \frac{2\pi\sqrt{2.4}}{0.633}$ (a,b) $b = 0.02$, $L = 8.4$, (c,d) $b = 1$ and $L = 1.68$}
    \label{fig5}
\end{figure}

Let us now turn to a more interesting and complex inhomogeneous slab which, nonetheless, has been found to posses an exact analytical solution for the scattered amplitudes \cite{jones2016extension}. A slab having PT symmetry at the symmetry-breaking point is described by $\xi(z) = be^{idz}$, with $b$ and $d$ positive numbers. The first and second-order Born approximations for the transmitted $t$ and reflected $r$ amplitudes are given by
\begin{equation}
\begin{split}
    t_1 &= \frac{kb}{2d\cos\theta} \left(e^{i L d} - 1 \right),\\
    t_2 &= \frac{k^2b^2}{8d\cos^2\theta} \left[\frac{d \left(e^{2 i L d}-2 e^{i L (d+2 k \cos \theta )}+1\right)}{(d-2k\cos\theta)(d+2k\cos\theta)}\right.\\
   &+\left. \frac{2 k
   \left(e^{2 i L d}-1\right) \cos \theta }{(d-2k\cos\theta)(d+2k\cos\theta)} + \frac{\left(e^{i L d} - 1\right)^2}{d}\right]
\end{split}
\end{equation}
\begin{equation}
\begin{split}
    r_1 &= \frac{kb}{2 \cos\theta}\left[\frac{e^{i L (d+2 k \cos \theta)} - 1}{d+2 k \cos\theta} \right],\\
    r_2 &= \frac{k^2b^2}{4 d \cos^2\theta} \left[\frac{e^{2 i L (d+k \cos \theta)}}{d+k \cos \theta  } - \frac{e^{2 i L (d+2k \cos\theta
   )}}{d+2 k \cos \theta }\right.\\
   &+\left. \frac{d}{(d+k \cos\theta) (d+2 k \cos \theta )}\right].
\end{split}
\end{equation}
We discuss only one aspect of this slab which is the behavior of the scattered amplitudes as functions of the incident angle $\theta$. By using the same parameters as in \cite{jones2016extension}, we obtain the Taylor series and construct the Pad\'e approximants. Parts (a) and (b) of Figure \ref{fig5} display the approximants $|t_2^2|^2$ and $|r_2^2|^2$. The exact solution is not shown along with the approximate one because both oscillate too fast and it would be very difficult to compare. We refer the reader to consult Figure 5 of \cite{jones2016extension}. 

With the particular numerical values taken from Ref. \cite{jones2016extension}, the first three terms in the Born series actually reproduce the exact behavior quite well. This is due to the fact that the amplitude of $\chi(z)$ is very small (0.02). Thus, we increase this amplitude in such a way that the Taylor series no longer converges and plot in parts (c) and (d) of the same figure, $|t_4^4|^2$ and $|r_4^4|^2$ with $b=1$. Again, the approximants converge and display an interesting behavior consisting of strong peaks in the transmission and corresponding troughs in the reflected amplitude. It is remarkable how the quotient of two polynomials can capture this rich dynamics. Note that only the first 9 terms of the Taylor series are used to construct the plots shown in parts (c) and (d).

\subsection{Bloch-type slab}

For this last example we consider a permittivity profile such that the Helmholtz equation has no known analytical exact solution. The profile $\xi(z) = \Gamma z + \sigma\cos(dz)$ has the form of a linear ramp superposed with a periodic modulation ($\sigma$ and $\Gamma$ are constants). It resembles the potential function of Schrodinger's equation that generates Bloch oscillations.

Since in this case we have no grounds for comparison with exact solutions, we choose the normalization condition $|t|^2+|r|^2 = 1$ to guarantee that the scattered amplitudes remain bounded. The first Born approximation for the transmitted and reflected amplitudes are given by
\begin{equation}
\begin{split}
    &t_1 = \frac{1}{2}ik\sec\theta \left[ \frac{\Gamma L^2}{2} + \frac{\sigma\sin(Ld)}{d}\right],\\
    &r_1 = \frac{1}{2}ik\sec\theta\left\{\frac{\Gamma \sec^2\theta\left[(1-2ikL\cos\theta)e^{2ikL\cos\theta}-1\right]}{4k^2}\right.\\
    & +\left. \frac{[d\sin(Ld) + 2ik\cos(Ld)\cos\theta]\sigma e^{2ikL\cos\theta}-2i\sigma k \cos\theta}{d^2-4k^2\cos^2\theta}\right\}.
\end{split}
\end{equation}

\begin{figure}[!htp]
    \centering
    \includegraphics[scale=0.67]{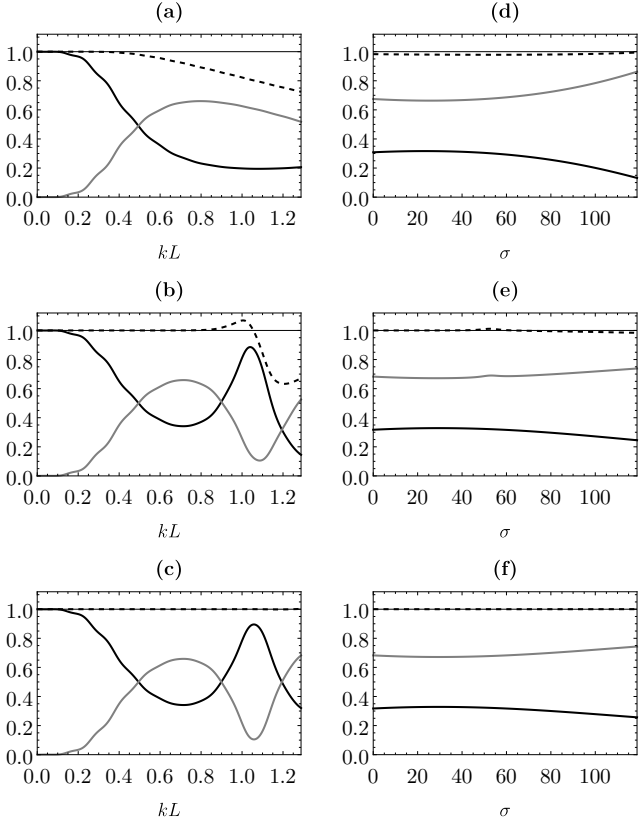}
   \caption{Transmitted and reflected amplitudes for the inhomogeneous slab $\xi (z) = \Gamma z + \sigma\cos(dz)$ as a function of (a-c) $kL$ and (d-f) $\sigma$. The continuous thick black line (continuous gray line) are the Pad\'e approximants for transmission $|t_N^N|^2$ (reflection $|r_N^N|^2$). The dashed line is the total energy. (a) and (d) $N=1$ (b) and (e) $N=2$ and (c) and (f) $N=3$. Parameters used: In all plots: $\alpha = 1$, $d = 50$ and $\Gamma = 20$. For (a-c), $\sigma=2$, $\theta = 0$ and for (d-f) $kL = 1/2$, $\theta = \pi/4 $. The corresponding Born series diverges for these set of parameters. The continuous thin black line marks the unitary value.}
    \label{fig6}
\end{figure}

Figure \ref{fig6}(a-c) shows the plot for the amplitudes as functions of $kL$. They resemble the amplitudes for the linear slab. However, there are small lumps present in the plot that do not appear in the linear case. This behavior arises from the oscillatory nature of the material. Parts (d-f) plot the scattered amplitudes as functions of $\sigma$ and we see once again the remarkable convergence properties of the approximants. In all cases the energy is conserved for sufficient large values of $N$.

\section{Conclusions}

We demonstrate the usefulness and importance of Pad\'e approximants in problems involving scattering of plane waves by inhomogeneous complex media. By using regular perturbation theory to obtain the reflected and transmitted amplitudes, divergent series are obtained which can be summed by constructing the approximants. The formalism is valid for arbitrary inhomogeneous materials and can handle oblique incidence.

\appendix

\section{Exact solution for a homogeneous slab}

In this appendix we provide the exact solution to the field amplitudes in the scattering system composed of a slab with size $L$ and homogeneous dielectric permittivity $\xi(z) = b$. The total fields in the region $z \leq 0$, $0 \leq z \leq L$ and $z \geq L$ are given by $\psi_1 = e^{ikz\cos\theta} + re^{-ikz\cos\theta}$, $\psi_2 = d_1 e^{ikz\sqrt{\cos^2\theta + \alpha b}} + d_2 e^{-ikz\sqrt{\cos^2\theta + \alpha b}}$ and $\psi_3 = te^{ikz\cos\theta}$, respectively. To connect the parameters $r$, $t$, $d_1$ and $d_2$ one uses the fact that the total field amplitude $\psi(z)$ and its derivative $d\psi(z)/dz$ are continuous at $z = 0$ and $z = L$. The first continuity condition comes directly from the continuity of the electric field parallel to the interface. The second boundary condition is a direct consequence of the differential equation satisfied by $\psi(z)$, or it can be viewed as a continuity condition for the magnetic field.

After straightforward algebra, the following system of equations is obtained
\begin{equation}\label{systemslab}
    \begin{split}
        &1 + r = d_1 + d_2, \\
        &d_1e^{ik \gamma L} + d_2e^{-ik\gamma L} = te^{ikL\cos\theta}, \\
        &(1 - r)\cos\theta = \gamma(d_1 - d_2), \\
        &\gamma\Big( d_1 e^{ik\gamma L} - d_2e^{-ik\gamma L} \Big) = t\cos\theta e^{ikL\cos\theta},
    \end{split}
\end{equation}
where $\gamma = \sqrt{\cos^2\theta + \alpha b}$. By solving the system we obtain the amplitudes
\begin{equation}
    d_1 = \frac{2(\gamma + \cos\theta)\cos\theta}{(\gamma + \cos\theta)^2 - (\gamma - \cos\theta)^2e^{2ik\gamma L}},
\end{equation}
\begin{equation}
    d_2 = \frac{2(\cos\theta - \gamma)\cos\theta}{(\gamma - \cos\theta)^2 - (\gamma + \cos\theta)^2e^{-2ik\gamma L}}.
\end{equation}
From the first (second) equation in \eqref{systemslab} we obtain the reflected (transmitted) amplitude $r$ ($t$).

\section{Exact solution for a linear slab}

In the case where $\xi(z) = 1 + bz$, the exact solution for the field amplitude $\psi_2(z)$ inside the slab is given by
\begin{equation}
    \psi_2(z) = c_1 \text{Ai}(u) + c_2\text{Bi}(u),
\end{equation}
where $c_1$ and $c_2$ are constants, $$u = \frac{-bk^2 z \alpha - \frac{k^2}{2}(1 + 2\alpha + \cos 2\theta)}{(-b\alpha k^2)^{2/3}}$$ and Ai and Bi are Airy functions defined as the linearly independent solutions of Airy's equation,
\begin{equation}
\begin{split}
    \text{Ai}(u) &= \sum_{n = 0}^{\infty} \frac{3^{\frac{n-2}{3}}}{\pi n!}\Gamma\left( \frac{n+1}{3} \right)\\
    & \qquad\qquad\times\sin\left[ \frac{2\pi}{3}(n+1) \right]u^n,\\
    \text{Bi}(u) &= \sum_{n = 0}^{\infty} \frac{3^{\frac{n-2}{3}}}{\pi n!}\Gamma\left( \frac{n+1}{3} \right) \\
    &\qquad\qquad\times\left\{\ 1 + \sin\left[ \frac{\pi}{6}(4n+1) \right]\right\}\ u^n,
    \end{split}
\end{equation}
where $\Gamma$ is the Gamma function. Both series converge in the finite complex plane because the differential equation has no singular points. By matching the free-space solutions for $z \leq 0$ and $z \geq L$ at both interfaces one obtains the transmitted and reflected amplitudes (written in terms of Airy functions and their derivatives).

\bibliography{refs}

\begin{thebibliography}{26}%
\makeatletter
\providecommand \@ifxundefined [1]{%
 \@ifx{#1\undefined}
}%
\providecommand \@ifnum [1]{%
 \ifnum #1\expandafter \@firstoftwo
 \else \expandafter \@secondoftwo
 \fi
}%
\providecommand \@ifx [1]{%
 \ifx #1\expandafter \@firstoftwo
 \else \expandafter \@secondoftwo
 \fi
}%
\providecommand \natexlab [1]{#1}%
\providecommand \enquote  [1]{``#1''}%
\providecommand \bibnamefont  [1]{#1}%
\providecommand \bibfnamefont [1]{#1}%
\providecommand \citenamefont [1]{#1}%
\providecommand \href@noop [0]{\@secondoftwo}%
\providecommand \href [0]{\begingroup \@sanitize@url \@href}%
\providecommand \@href[1]{\@@startlink{#1}\@@href}%
\providecommand \@@href[1]{\endgroup#1\@@endlink}%
\providecommand \@sanitize@url [0]{\catcode `\\12\catcode `\$12\catcode
  `\&12\catcode `\#12\catcode `\^12\catcode `\_12\catcode `\%12\relax}%
\providecommand \@@startlink[1]{}%
\providecommand \@@endlink[0]{}%
\providecommand \url  [0]{\begingroup\@sanitize@url \@url }%
\providecommand \@url [1]{\endgroup\@href {#1}{\urlprefix }}%
\providecommand \urlprefix  [0]{URL }%
\providecommand \Eprint [0]{\href }%
\providecommand \doibase [0]{https://doi.org/}%
\providecommand \selectlanguage [0]{\@gobble}%
\providecommand \bibinfo  [0]{\@secondoftwo}%
\providecommand \bibfield  [0]{\@secondoftwo}%
\providecommand \translation [1]{[#1]}%
\providecommand \BibitemOpen [0]{}%
\providecommand \bibitemStop [0]{}%
\providecommand \bibitemNoStop [0]{.\EOS\space}%
\providecommand \EOS [0]{\spacefactor3000\relax}%
\providecommand \BibitemShut  [1]{\csname bibitem#1\endcsname}%
\let\auto@bib@innerbib\@empty
\bibitem [{\citenamefont {Born}\ and\ \citenamefont
  {Wolf}(2013)}]{born2013principles}%
  \BibitemOpen
  \bibfield  {author} {\bibinfo {author} {\bibfnamefont {M.}~\bibnamefont
  {Born}}\ and\ \bibinfo {author} {\bibfnamefont {E.}~\bibnamefont {Wolf}},\
  }\href@noop {} {\emph {\bibinfo {title} {Principles of optics:
  electromagnetic theory of propagation, interference and diffraction of
  light}}}\ (\bibinfo  {publisher} {Elsevier},\ \bibinfo {year}
  {2013})\BibitemShut {NoStop}%
\bibitem [{\citenamefont {Saleh}\ and\ \citenamefont
  {Teich}(2019)}]{saleh2019fundamentals}%
  \BibitemOpen
  \bibfield  {author} {\bibinfo {author} {\bibfnamefont {B.~E.}\ \bibnamefont
  {Saleh}}\ and\ \bibinfo {author} {\bibfnamefont {M.~C.}\ \bibnamefont
  {Teich}},\ }\href@noop {} {\emph {\bibinfo {title} {Fundamentals of
  photonics}}}\ (\bibinfo  {publisher} {john Wiley \& sons},\ \bibinfo {year}
  {2019})\BibitemShut {NoStop}%
\bibitem [{\citenamefont {Bender}\ \emph {et~al.}(1999)\citenamefont {Bender},
  \citenamefont {Orszag},\ and\ \citenamefont {Orszag}}]{bender1999advanced}%
  \BibitemOpen
  \bibfield  {author} {\bibinfo {author} {\bibfnamefont {C.~M.}\ \bibnamefont
  {Bender}}, \bibinfo {author} {\bibfnamefont {S.}~\bibnamefont {Orszag}},\
  and\ \bibinfo {author} {\bibfnamefont {S.~A.}\ \bibnamefont {Orszag}},\
  }\href@noop {} {\emph {\bibinfo {title} {Advanced mathematical methods for
  scientists and engineers I: Asymptotic methods and perturbation theory}}},\
  Vol.~\bibinfo {volume} {1}\ (\bibinfo  {publisher} {Springer Science \&
  Business Media},\ \bibinfo {year} {1999})\BibitemShut {NoStop}%
\bibitem [{\citenamefont {Bremmer}(1951)}]{bremmer1951wkb}%
  \BibitemOpen
  \bibfield  {author} {\bibinfo {author} {\bibfnamefont {H.}~\bibnamefont
  {Bremmer}},\ }\bibfield  {title} {\bibinfo {title} {The wkb approximation as
  the first term of a geometric-optical series},\ }\href@noop {} {\bibfield
  {journal} {\bibinfo  {journal} {Communications on pure and applied
  mathematics}\ }\textbf {\bibinfo {volume} {4}},\ \bibinfo {pages} {105}
  (\bibinfo {year} {1951})}\BibitemShut {NoStop}%
\bibitem [{\citenamefont {Atkinson}(1960)}]{atkinson1960wave}%
  \BibitemOpen
  \bibfield  {author} {\bibinfo {author} {\bibfnamefont {F.}~\bibnamefont
  {Atkinson}},\ }\bibfield  {title} {\bibinfo {title} {Wave propagation and the
  bremmer series},\ }\href@noop {} {\bibfield  {journal} {\bibinfo  {journal}
  {Journal of mathematical analysis and applications}\ }\textbf {\bibinfo
  {volume} {1}},\ \bibinfo {pages} {255} (\bibinfo {year} {1960})}\BibitemShut
  {NoStop}%
\bibitem [{\citenamefont {Berk}\ \emph {et~al.}(1967)\citenamefont {Berk},
  \citenamefont {Book},\ and\ \citenamefont {Pfirsch}}]{berk1967convergence}%
  \BibitemOpen
  \bibfield  {author} {\bibinfo {author} {\bibfnamefont {H.}~\bibnamefont
  {Berk}}, \bibinfo {author} {\bibfnamefont {D.}~\bibnamefont {Book}},\ and\
  \bibinfo {author} {\bibfnamefont {D.}~\bibnamefont {Pfirsch}},\ }\bibfield
  {title} {\bibinfo {title} {Convergence of the bremmer series for the
  spatially inhomogeneous helmholtz equation},\ }\href@noop {} {\bibfield
  {journal} {\bibinfo  {journal} {Journal of Mathematical Physics}\ }\textbf
  {\bibinfo {volume} {8}},\ \bibinfo {pages} {1611} (\bibinfo {year}
  {1967})}\BibitemShut {NoStop}%
\bibitem [{\citenamefont {Hassab}(1972)}]{hassab1972perturbational}%
  \BibitemOpen
  \bibfield  {author} {\bibinfo {author} {\bibfnamefont {J.}~\bibnamefont
  {Hassab}},\ }\bibfield  {title} {\bibinfo {title} {Perturbational solution of
  the helmholtz equation in arbitrary inhomogeneous media},\ }\href@noop {}
  {\bibfield  {journal} {\bibinfo  {journal} {IEEE Transactions on Antennas and
  Propagation}\ }\textbf {\bibinfo {volume} {20}},\ \bibinfo {pages} {524}
  (\bibinfo {year} {1972})}\BibitemShut {NoStop}%
\bibitem [{\citenamefont {Chen}(1978)}]{chen1978integral}%
  \BibitemOpen
  \bibfield  {author} {\bibinfo {author} {\bibfnamefont {C.}~\bibnamefont
  {Chen}},\ }\bibfield  {title} {\bibinfo {title} {An integral equation
  formulation of the direct scattering problem for an inhomogeneous slab},\
  }\href@noop {} {\bibfield  {journal} {\bibinfo  {journal} {IEEE Transactions
  on Antennas and Propagation}\ }\textbf {\bibinfo {volume} {26}},\ \bibinfo
  {pages} {797} (\bibinfo {year} {1978})}\BibitemShut {NoStop}%
\bibitem [{\citenamefont {Chen}\ and\ \citenamefont
  {Lien}(1979)}]{chen1979finite}%
  \BibitemOpen
  \bibfield  {author} {\bibinfo {author} {\bibfnamefont {C.}~\bibnamefont
  {Chen}}\ and\ \bibinfo {author} {\bibfnamefont {C.-D.}\ \bibnamefont
  {Lien}},\ }\bibfield  {title} {\bibinfo {title} {A finite element solution of
  the wave propagation problem for an inhomogeneous dielectric slab},\
  }\href@noop {} {\bibfield  {journal} {\bibinfo  {journal} {IEEE Transactions
  on Antennas and Propagation}\ }\textbf {\bibinfo {volume} {27}},\ \bibinfo
  {pages} {877} (\bibinfo {year} {1979})}\BibitemShut {NoStop}%
\bibitem [{\citenamefont {Chen}\ and\ \citenamefont
  {Kiang}(1980)}]{chen1980variational}%
  \BibitemOpen
  \bibfield  {author} {\bibinfo {author} {\bibfnamefont {C.-H.}\ \bibnamefont
  {Chen}}\ and\ \bibinfo {author} {\bibfnamefont {Y.-W.}\ \bibnamefont
  {Kiang}},\ }\bibfield  {title} {\bibinfo {title} {A variational theory for
  wave propagation in a one-dimensional inhomogeneous medium},\ }\href@noop {}
  {\bibfield  {journal} {\bibinfo  {journal} {IEEE Transactions on Antennas and
  Propagation}\ }\textbf {\bibinfo {volume} {28}},\ \bibinfo {pages} {762}
  (\bibinfo {year} {1980})}\BibitemShut {NoStop}%
\bibitem [{\citenamefont {Su}\ and\ \citenamefont {Chen}(1982)}]{su1982fast}%
  \BibitemOpen
  \bibfield  {author} {\bibinfo {author} {\bibfnamefont {C.-C.}\ \bibnamefont
  {Su}}\ and\ \bibinfo {author} {\bibfnamefont {C.-H.}\ \bibnamefont {Chen}},\
  }\bibfield  {title} {\bibinfo {title} {A fast algorithm for inhomogeneous
  slab scattering problems from the integral equation approach},\ }\href@noop
  {} {\bibfield  {journal} {\bibinfo  {journal} {Journal of Applied Physics}\
  }\textbf {\bibinfo {volume} {53}},\ \bibinfo {pages} {6009} (\bibinfo {year}
  {1982})}\BibitemShut {NoStop}%
\bibitem [{\citenamefont {Longhi}(2018)}]{longhi2018parity}%
  \BibitemOpen
  \bibfield  {author} {\bibinfo {author} {\bibfnamefont {S.}~\bibnamefont
  {Longhi}},\ }\bibfield  {title} {\bibinfo {title} {Parity-time symmetry meets
  photonics: A new twist in non-hermitian optics},\ }\href@noop {} {\bibfield
  {journal} {\bibinfo  {journal} {EPL (Europhysics Letters)}\ }\textbf
  {\bibinfo {volume} {120}},\ \bibinfo {pages} {64001} (\bibinfo {year}
  {2018})}\BibitemShut {NoStop}%
\bibitem [{\citenamefont {El-Ganainy}\ \emph {et~al.}(2019)\citenamefont
  {El-Ganainy}, \citenamefont {Khajavikhan}, \citenamefont {Christodoulides},\
  and\ \citenamefont {Ozdemir}}]{el2019dawn}%
  \BibitemOpen
  \bibfield  {author} {\bibinfo {author} {\bibfnamefont {R.}~\bibnamefont
  {El-Ganainy}}, \bibinfo {author} {\bibfnamefont {M.}~\bibnamefont
  {Khajavikhan}}, \bibinfo {author} {\bibfnamefont {D.~N.}\ \bibnamefont
  {Christodoulides}},\ and\ \bibinfo {author} {\bibfnamefont {S.~K.}\
  \bibnamefont {Ozdemir}},\ }\bibfield  {title} {\bibinfo {title} {The dawn of
  non-hermitian optics},\ }\href@noop {} {\bibfield  {journal} {\bibinfo
  {journal} {Communications Physics}\ }\textbf {\bibinfo {volume} {2}},\
  \bibinfo {pages} {1} (\bibinfo {year} {2019})}\BibitemShut {NoStop}%
\bibitem [{\citenamefont {Feng}\ \emph {et~al.}(2017)\citenamefont {Feng},
  \citenamefont {El-Ganainy},\ and\ \citenamefont {Ge}}]{feng2017non}%
  \BibitemOpen
  \bibfield  {author} {\bibinfo {author} {\bibfnamefont {L.}~\bibnamefont
  {Feng}}, \bibinfo {author} {\bibfnamefont {R.}~\bibnamefont {El-Ganainy}},\
  and\ \bibinfo {author} {\bibfnamefont {L.}~\bibnamefont {Ge}},\ }\bibfield
  {title} {\bibinfo {title} {Non-hermitian photonics based on parity--time
  symmetry},\ }\href@noop {} {\bibfield  {journal} {\bibinfo  {journal} {Nature
  Photonics}\ }\textbf {\bibinfo {volume} {11}},\ \bibinfo {pages} {752}
  (\bibinfo {year} {2017})}\BibitemShut {NoStop}%
\bibitem [{\citenamefont {Bender}\ and\ \citenamefont
  {Boettcher}(1998)}]{bender1998real}%
  \BibitemOpen
  \bibfield  {author} {\bibinfo {author} {\bibfnamefont {C.~M.}\ \bibnamefont
  {Bender}}\ and\ \bibinfo {author} {\bibfnamefont {S.}~\bibnamefont
  {Boettcher}},\ }\bibfield  {title} {\bibinfo {title} {Real spectra in
  non-hermitian hamiltonians having p t symmetry},\ }\href@noop {} {\bibfield
  {journal} {\bibinfo  {journal} {Physical review letters}\ }\textbf {\bibinfo
  {volume} {80}},\ \bibinfo {pages} {5243} (\bibinfo {year}
  {1998})}\BibitemShut {NoStop}%
\bibitem [{\citenamefont {Ge}\ \emph {et~al.}(2011)\citenamefont {Ge},
  \citenamefont {Chong}, \citenamefont {Rotter}, \citenamefont {T{\"u}reci},\
  and\ \citenamefont {Stone}}]{ge2011unconventional}%
  \BibitemOpen
  \bibfield  {author} {\bibinfo {author} {\bibfnamefont {L.}~\bibnamefont
  {Ge}}, \bibinfo {author} {\bibfnamefont {Y.}~\bibnamefont {Chong}}, \bibinfo
  {author} {\bibfnamefont {S.}~\bibnamefont {Rotter}}, \bibinfo {author}
  {\bibfnamefont {H.~E.}\ \bibnamefont {T{\"u}reci}},\ and\ \bibinfo {author}
  {\bibfnamefont {A.}~\bibnamefont {Stone}},\ }\bibfield  {title} {\bibinfo
  {title} {Unconventional modes in lasers with spatially varying gain and
  loss},\ }\href@noop {} {\bibfield  {journal} {\bibinfo  {journal} {Physical
  Review A}\ }\textbf {\bibinfo {volume} {84}},\ \bibinfo {pages} {023820}
  (\bibinfo {year} {2011})}\BibitemShut {NoStop}%
\bibitem [{\citenamefont {Lin}\ \emph {et~al.}(2011)\citenamefont {Lin},
  \citenamefont {Ramezani}, \citenamefont {Eichelkraut}, \citenamefont
  {Kottos}, \citenamefont {Cao},\ and\ \citenamefont
  {Christodoulides}}]{lin2011unidirectional}%
  \BibitemOpen
  \bibfield  {author} {\bibinfo {author} {\bibfnamefont {Z.}~\bibnamefont
  {Lin}}, \bibinfo {author} {\bibfnamefont {H.}~\bibnamefont {Ramezani}},
  \bibinfo {author} {\bibfnamefont {T.}~\bibnamefont {Eichelkraut}}, \bibinfo
  {author} {\bibfnamefont {T.}~\bibnamefont {Kottos}}, \bibinfo {author}
  {\bibfnamefont {H.}~\bibnamefont {Cao}},\ and\ \bibinfo {author}
  {\bibfnamefont {D.~N.}\ \bibnamefont {Christodoulides}},\ }\bibfield  {title}
  {\bibinfo {title} {Unidirectional invisibility induced by p t-symmetric
  periodic structures},\ }\href@noop {} {\bibfield  {journal} {\bibinfo
  {journal} {Physical Review Letters}\ }\textbf {\bibinfo {volume} {106}},\
  \bibinfo {pages} {213901} (\bibinfo {year} {2011})}\BibitemShut {NoStop}%
\bibitem [{\citenamefont {Jones}(2012)}]{jones2012analytic}%
  \BibitemOpen
  \bibfield  {author} {\bibinfo {author} {\bibfnamefont {H.}~\bibnamefont
  {Jones}},\ }\bibfield  {title} {\bibinfo {title} {Analytic results for a
  pt-symmetric optical structure},\ }\href@noop {} {\bibfield  {journal}
  {\bibinfo  {journal} {Journal of Physics A: Mathematical and Theoretical}\
  }\textbf {\bibinfo {volume} {45}},\ \bibinfo {pages} {135306} (\bibinfo
  {year} {2012})}\BibitemShut {NoStop}%
\bibitem [{\citenamefont {Kalish}\ \emph {et~al.}(2012)\citenamefont {Kalish},
  \citenamefont {Lin},\ and\ \citenamefont {Kottos}}]{kalish2012light}%
  \BibitemOpen
  \bibfield  {author} {\bibinfo {author} {\bibfnamefont {S.}~\bibnamefont
  {Kalish}}, \bibinfo {author} {\bibfnamefont {Z.}~\bibnamefont {Lin}},\ and\
  \bibinfo {author} {\bibfnamefont {T.}~\bibnamefont {Kottos}},\ }\bibfield
  {title} {\bibinfo {title} {Light transport in random media with pt
  symmetry},\ }\href@noop {} {\bibfield  {journal} {\bibinfo  {journal}
  {Physical Review A}\ }\textbf {\bibinfo {volume} {85}},\ \bibinfo {pages}
  {055802} (\bibinfo {year} {2012})}\BibitemShut {NoStop}%
\bibitem [{\citenamefont {Zhu}\ \emph {et~al.}(2014)\citenamefont {Zhu},
  \citenamefont {Peng},\ and\ \citenamefont {Zhao}}]{zhu2014anisotropic}%
  \BibitemOpen
  \bibfield  {author} {\bibinfo {author} {\bibfnamefont {X.-F.}\ \bibnamefont
  {Zhu}}, \bibinfo {author} {\bibfnamefont {Y.-G.}\ \bibnamefont {Peng}},\ and\
  \bibinfo {author} {\bibfnamefont {D.-G.}\ \bibnamefont {Zhao}},\ }\bibfield
  {title} {\bibinfo {title} {Anisotropic reflection oscillation in periodic
  multilayer structures of parity-time symmetry},\ }\href@noop {} {\bibfield
  {journal} {\bibinfo  {journal} {Optics express}\ }\textbf {\bibinfo {volume}
  {22}},\ \bibinfo {pages} {18401} (\bibinfo {year} {2014})}\BibitemShut
  {NoStop}%
\bibitem [{\citenamefont {Baker}\ \emph {et~al.}(1996)\citenamefont {Baker},
  \citenamefont {Baker~Jr}, \citenamefont {Graves-Morris}, \citenamefont
  {Baker},\ and\ \citenamefont {Baker}}]{baker1996pade}%
  \BibitemOpen
  \bibfield  {author} {\bibinfo {author} {\bibfnamefont {G.~A.}\ \bibnamefont
  {Baker}}, \bibinfo {author} {\bibfnamefont {G.~A.}\ \bibnamefont {Baker~Jr}},
  \bibinfo {author} {\bibfnamefont {P.}~\bibnamefont {Graves-Morris}}, \bibinfo
  {author} {\bibfnamefont {G.}~\bibnamefont {Baker}},\ and\ \bibinfo {author}
  {\bibfnamefont {S.~S.}\ \bibnamefont {Baker}},\ }\href@noop {} {\emph
  {\bibinfo {title} {Pade Approximants: Encyclopedia of Mathematics and It's
  Applications, Vol. 59 George A. Baker, Jr., Peter Graves-Morris}}},\
  Vol.~\bibinfo {volume} {59}\ (\bibinfo  {publisher} {Cambridge University
  Press},\ \bibinfo {year} {1996})\BibitemShut {NoStop}%
\bibitem [{\citenamefont {Brezinski}(2012)}]{brezinski2012history}%
  \BibitemOpen
  \bibfield  {author} {\bibinfo {author} {\bibfnamefont {C.}~\bibnamefont
  {Brezinski}},\ }\href@noop {} {\emph {\bibinfo {title} {History of continued
  fractions and Pad{\'e} approximants}}},\ Vol.~\bibinfo {volume} {12}\
  (\bibinfo  {publisher} {Springer Science \& Business Media},\ \bibinfo {year}
  {2012})\BibitemShut {NoStop}%
\bibitem [{\citenamefont {George~Jr}\ \emph {et~al.}(1975)\citenamefont
  {George~Jr} \emph {et~al.}}]{george1975essentials}%
  \BibitemOpen
  \bibfield  {author} {\bibinfo {author} {\bibfnamefont {A.}~\bibnamefont
  {George~Jr}} \emph {et~al.},\ }\href@noop {} {\emph {\bibinfo {title}
  {Essentials of Pad{\'e} approximants}}}\ (\bibinfo  {publisher} {Elsevier},\
  \bibinfo {year} {1975})\BibitemShut {NoStop}%
\bibitem [{\citenamefont {Rebou{\c{c}}as}\ and\ \citenamefont
  {Brand{\~a}o}(2021)}]{rebouccas2021scattering}%
  \BibitemOpen
  \bibfield  {author} {\bibinfo {author} {\bibfnamefont {J.~A.}\ \bibnamefont
  {Rebou{\c{c}}as}}\ and\ \bibinfo {author} {\bibfnamefont {P.~A.}\
  \bibnamefont {Brand{\~a}o}},\ }\bibfield  {title} {\bibinfo {title}
  {Scattering of light by a parity-time-symmetric dipole beyond the first born
  approximation},\ }\href@noop {} {\bibfield  {journal} {\bibinfo  {journal}
  {Physical Review A}\ }\textbf {\bibinfo {volume} {104}},\ \bibinfo {pages}
  {063514} (\bibinfo {year} {2021})}\BibitemShut {NoStop}%
\bibitem [{\citenamefont {van~der Sijs}\ \emph {et~al.}(2020)\citenamefont
  {van~der Sijs}, \citenamefont {El~Gawhary},\ and\ \citenamefont
  {Urbach}}]{van2020electromagnetic}%
  \BibitemOpen
  \bibfield  {author} {\bibinfo {author} {\bibfnamefont {T.}~\bibnamefont
  {van~der Sijs}}, \bibinfo {author} {\bibfnamefont {O.}~\bibnamefont
  {El~Gawhary}},\ and\ \bibinfo {author} {\bibfnamefont {H.}~\bibnamefont
  {Urbach}},\ }\bibfield  {title} {\bibinfo {title} {Electromagnetic scattering
  beyond the weak regime: Solving the problem of divergent born perturbation
  series by pad{\'e} approximants},\ }\href@noop {} {\bibfield  {journal}
  {\bibinfo  {journal} {Physical Review Research}\ }\textbf {\bibinfo {volume}
  {2}},\ \bibinfo {pages} {013308} (\bibinfo {year} {2020})}\BibitemShut
  {NoStop}%
\bibitem [{\citenamefont {Jones}\ and\ \citenamefont
  {Kulishov}(2016)}]{jones2016extension}%
  \BibitemOpen
  \bibfield  {author} {\bibinfo {author} {\bibfnamefont {H.}~\bibnamefont
  {Jones}}\ and\ \bibinfo {author} {\bibfnamefont {M.}~\bibnamefont
  {Kulishov}},\ }\bibfield  {title} {\bibinfo {title} {Extension of analytic
  results for a pt-symmetric structure},\ }\href@noop {} {\bibfield  {journal}
  {\bibinfo  {journal} {Journal of Optics}\ }\textbf {\bibinfo {volume} {18}},\
  \bibinfo {pages} {055101} (\bibinfo {year} {2016})}\BibitemShut {NoStop}%
\end{thebibliography}%

\end{document}